# Anomalous behavior of the quasi-one-dimensional quantum material Na$_2$OsO$_4$ at high pressure

Raimundas Sereika [1,*], Kazunari Yamaura [2], Yating Jia [3], Sijia Zhang [3], Changqing Jin [3,4,5], Hongkee Yoon [6], Min Yong Jeong [6], Myung Joon Han [6,7], Dale L. Brewe [8], Steve M. Heald [8], Stanislav Sinogeikin [9,10], Yang Ding [1,*], Ho-kwang Mao [1,10,11]

[1] *Center for High Pressure Science and Technology Advanced Research, Beijing 100094, China*

[2] *National Institute for Materials Science, 1-1 Namiki, Tsukuba, Ibaraki 305-0044, Japan*

[3] *Institute of Physics, Chinese Academy of Sciences, Beijing 100190, China*

[4] *School of Physical Science, University of Chinese Academy of Sciences, Beijing 100190, China*

[5] *Collaborative Innovation Center of Quantum Matter, Beijing 100084, China*

[6] *Department of Physics, KAIST, 291 Daehak-ro, Yuseong-gu, Daejeon 34141, Korea*

[7] *KAIST Institute for the NanoCentury, Korea Advanced Institute of Science and Technology, Daejeon 305-701, Korea*

[8] *X-ray Science Division, Advanced Photon Source, Argonne National Laboratory, 9700 South Cass Avenue, Lemont, Illinois 60439, USA*

[9] *DACTools LLC, Naperville, Illinois 60565, USA*

[10] *HPCAT, Geophysical Laboratory, Carnegie Institution of Washington, 9700 South Cass Avenue, Lemont, Illinois 60439, USA*

[11] *Geophysical Laboratory, Carnegie Institution of Washington, Washington DC 20015, USA.*

*Corresponding authors:
raimundas.sereika@hpstar.ac.cn;
yang.ding@hpstar.ac.cn*

**Abstract**

Na$_2$OsO$_4$ is an unusual quantum material that, in contrast to the common 5$d^2$ oxides with spins = 1, owns a magnetically silent ground state with spin = 0 and a band gap at Fermi level attributed to a distortion in the OsO$_6$ octahedral sites. In this semiconductor, our low-temperature electrical transport measurements indicate an anomaly at 6.3 K with a power-law behavior inclining through the semiconductor-to-metal transition observed at 23 GPa. Even more peculiarly, we discover that before this transition, the material becomes more insulating instead of merely turning into a metal according to the conventional wisdom. To investigate the underlying mechanisms, we applied experimental and theoretical methods to examine the electronic and crystal structures comprehensively, and conclude that the enhanced insulating state at high pressure originates from the enlarged distortion of the OsO$_6$. It is such a distortion that widens the band gap and decreases the electron occupancy in Os's $t_{2g}$ orbital through an interplay of the lattice, charge, and orbital in the material, which is responsible for the changes observed in our experiments.

The structure of $Na_2OsO_4$ is characteristic of quasi-one-dimensional (quasi-1-D) anisotropy, reflecting a notable chain structure in which each chain comprises of edge-shared $OsO_6$ octahedra [1,2]. In such a system, *Fermi* liquid (FL) theory no longer applies because of the strongly correlated electronic behavior confined in the narrow channels [3,4]. Thus, the properties initiated from the collective behaviors of particles may appear unlike the effects from constituent individuals. Having a hexagonal lattice, $Na_2OsO_4$ is comparable to stoichiometrically equivalent $Ca_2IrO_4$ [5], but the magnetic and electrical measurements revealed its electrically semiconducting and nonmagnetic behavior. This result notably contradicted expectations given that Os has a $5d^2$ electronic configuration, which usually leads to unpaired spin moments and some contributions from orbital moments. The absence of magnetic moments was examined experimentally, and a remarkable distortion of the $OsO_6$ octahedra was found, even though $Os^{6+}$ is not strongly Jahn–Teller active [2,6]. Thus, it became clear that the degeneracy of the $5d_{xy}$ and $5d_{yz}$ ($5d_{zx}$) orbitals is broken and $Na_2OsO_4$ is magnetically silent in its ground state due to spin pairing ($S = 0$). This fact was also supported using *ab-initio* electronic structure calculation [1]. The scientifically intriguing aspect is that the energy gap ($E_g$) in such a simple system is triggered by the axial compression of the $OsO_6$ octahedra with a negligible role of SOC.

Under normal conditions, in the nondistorted octahedra, the $5d$ orbitals are splitted into $e_g$ (with $d_{z^2}$, $d_{x^2-y^2}$ two orbitals being degenerate) and $t_{2g}$ (with $d_{xy}$, $d_{yz}$, and $d_{zx}$ three orbitals being degenerate). The two $5d$ electrons fill into $t_{2g}$ orbital with two parallel spins according to the Hund's rules. This configuration gives a total local spin = 1 for each $OsO_6$ octahedral site. However, the octahedral distortion further splits the $t_{2g}$ orbitals into three energy different levels $d_{xy}$, $d_{yz}$, $d_{zx}$ leaving two $5d$ electrons in the lowest level ($d_{xy}$) with a pair of anti-parallel spins. Correspondingly, the total local spin = 0, which is magnetically silent state for each $OsO_6$ octahedral site in $Na_2OsO_4$. At high pressure, once the $e_g$ and $t_{2g}$ orbitals become overlapped, the two $5d$ electrons are no longer confined in the $t_{2g}$ orbitals but become delocalized into the both $e_g$ and $t_{2g}$ orbitals, then the total local spin is decreased and form a band-like magnetism. Since the magnetically silent state was directly related to the $OsO_6$ octahedra, physical or chemical pressure may render a significant impact on the structure by removing the $OsO_6$ distortion that exists at normal conditions. During compression, narrow $E_g$ semiconductors usually switch to metals and the non-magnetic $S = 0$ state could transform to a magnetic $S = 1$ state since the $OsO_6$ distortion can be removed or suppressed by pressure [7,8]. To search for the predicted quantum state, we combined high-pressure electrical transport, Synchrotron X-ray absorption (XAS), and diffraction techniques (XRD) in line with first-principles calculations to study the electronic and crystal structure at high pressure comprehensively. Contrary to expectations, however, we discovered that $Na_2OsO_4$ becomes increasingly insulating up to the 11 GPa, and then gradually transforms into a metallic state at 23 GPa. According to our theoretical modeling, we concluded that the increased insulating state originates from the pressure-enhanced distortion of the $OsO_6$ octahedra up to 11 GPa, which is also responsible for the unusual changes observed in the electrical transport and XAS experiments.

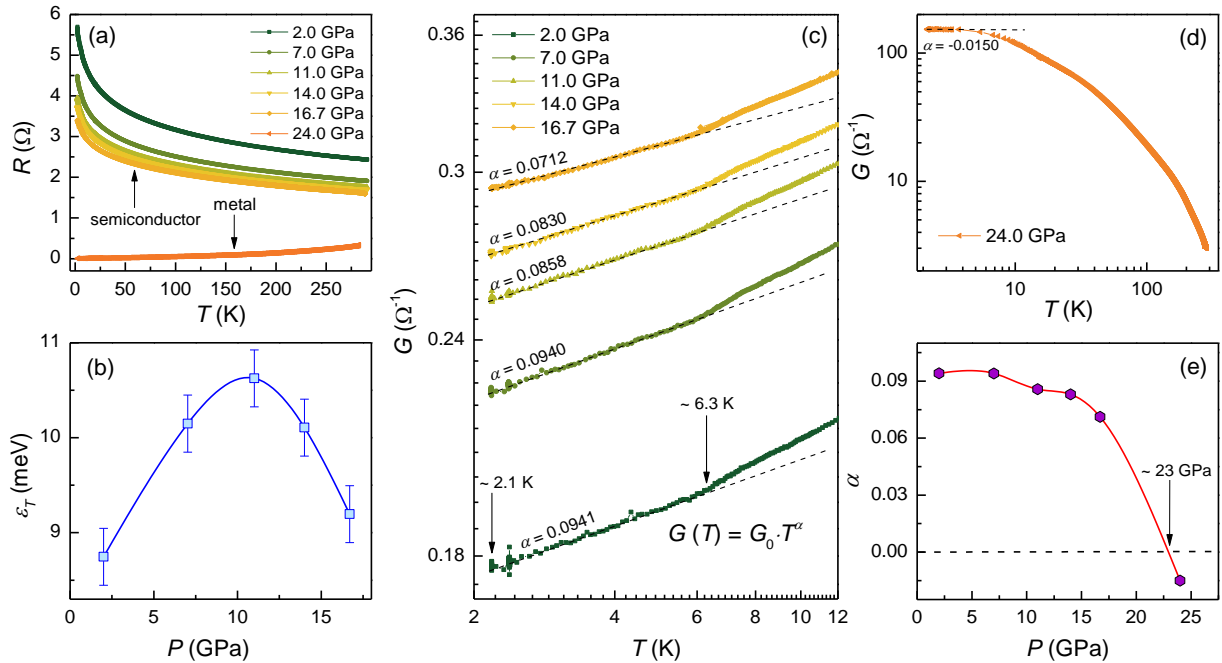

**Figure 1.** The temperature dependent electrical transport at fixed pressures. (a) Temperature dependence of the resistance (*R*) at various pressures. (b) The activation energy ($\varepsilon_T$) as a function of pressure (*P*) extracted from higher temperature region of the electrical transport. (c) and (d) represent log-log plots of the conductance (*G*) *versus* temperature for semiconducting and metallic states, respectively. The power-law behavior observed at low temperatures is indicated by dashed lines, together with an exponent parameter $\alpha$ at each pressure. (d) The exponent $\alpha$ dependence on pressure where the red curve is a guide to the eye. Here the 0 value is crossed at around 23 GPa.

Figure 1 shows the temperature dependent electrical transport properties of $Na_2OsO_4$. The $R(T)$ curves represent semiconducting behavior up to ~ 23 GPa (see Fig. 1a and data given in the Supporting Information). However, in this range, the activation energy ($\varepsilon_T$) deduced from the slope of the linear course of the electrical resistance for higher temperature region indicates non-linear curse (Fig. 1b and Fig. S4). The $\varepsilon_T(P)$ shows a maximum at 11 GPa, distinctly manifesting anomalous behavior at this pressure. Furthermore, at very low temperatures (from ~ 6.3 K), the conductance ($G$) exhibits clear power-law behavior, $G \propto T^\alpha$, which refer to the Luttinger liquid or 1-D Wigner crystal formation typical for 1-D systems (Figs. 1c, d). Therefore, we postulate that the observed phenomenon at low temperatures originates from the domain which in turn causes the tunneling effect among different Luttinger liquids.

It is worth mentioning that the power law does not extend to all temperature as pressure increases, as seen in some nanomaterials [9,10]. The obtained exponential values $\alpha$ were also relatively small ranging near 0 (see Figs. 1c, d). The compression rapidly decreased $\alpha$ following a polynomial trend where the power-law in metallic state (pressures > 23 GPa) appeared to be with an opposite sign (Fig. 1e). As mentioned above, the interacting fermions in one spatial dimension do not obey FL theory; however, the possibility of the deconfinement transition induced by interchain hopping [11] or a transition to a weakly disordered Fermi liquid [12] for more higher pressures cannot be neglected as $\alpha$ alters with pressure strongly towards FL state. Thus, such unique behavior at low temperature and high pressure can be further addressed by using more specific techniques.

For the underlying mechanisms at room temperature and specifically at 11 GPa, we carried out X-ray absorption near edge structure (XANES) measurements. Figures 2a and 2b show the Os $L_{3,2}$-edges captured at room temperature up to 28 GPa, respectively. The white-line (WL) energy position of $Na_2OsO_4$ Os $L_3$-edge corresponded well to the $5d^2$ samples [13,14] confirming the $Os^{6+}$ valence state at ambient pressure. Nevertheless, both the $L_3$ ($2p_{3/2} \rightarrow 5d$) and $L_2$ ($2p_{1/2} \rightarrow 5d$) absorption edges showed significant and unusual changes during compression. The intensity of the WL at the $L_3$ manifested a strong decrease during compression from 1 bar to 11.2 GPa and a strong increase from 11.2 GPa to 28.0 GPa (see Fig. 2c). The $L_2$ absorption edge did not indicate such a significant intensity change.

The changes under compression were also noticed in the Os WL position at the $L_3$ edge (Fig. 2c). The WL shift was observed in $Na_2OsO_4$ powders, which is also confirmed by measuring several single crystals with appropriate thicknesses. The WL shifts gradually to higher energies while under pressure up to 11.2 GPa and then reverts to its original position if more pressure is applied. As the WL position defines the valence, its shift to higher energies indicates its increase. A spectral shift at the Os $L_3$ edge of approximately 1.0(1) eV to higher energy indicates an increase of valence by 1+ state [14]. Based on this fact, we consider the increase of the Os valence (according to the WL position shift from powder data) from 6+ to 6.724+ during compression from 1 bar to 11.2 GPa, and vice versa, a decrease of Os valence from 6.724+ to 6+ during further compression from 11.2 GPa to 28 GPa. The increase and decrease of the Os valence might be associated with the $OsO_6$ octahedral deformation as it is known that there is an inverse relationship between the extent of the edge energy and the average bond length [16,17].

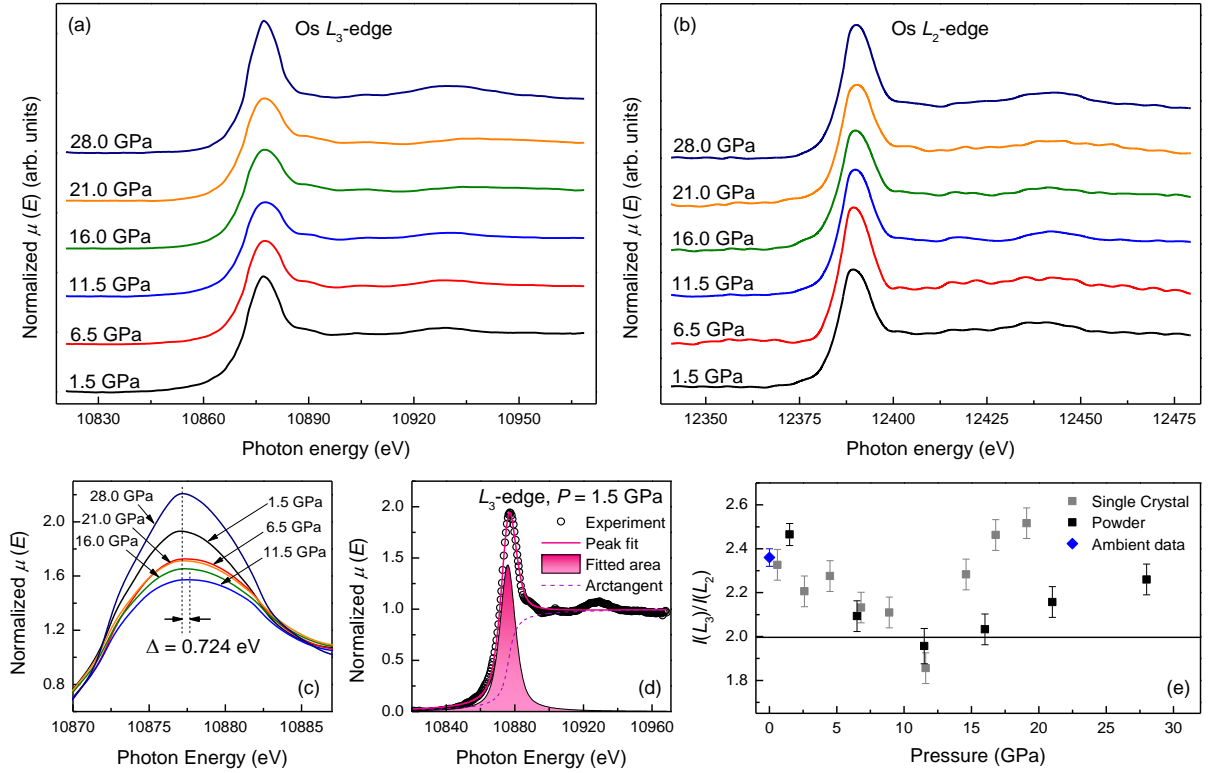

**Figure 2.** The XANES data of Os $L_3$ and $L_2$ - edges in $Na_2OsO_4$ at various pressures. (a) and (b) Evolution of the absorption spectra in a wide energy range during sample compression up to 28 GPa for the Os $L_3$ and $L_2$ – edges, respectively. (c) A comparison of the $L_3$-edge apex from (a) plot. (d) The determination of the white-line intensity at selected pressures. The open circles are the experimental results of the normalized $\mu(E)$ coefficient at the $L_3$ absorption edge, while the solid pink curve represents the best fit to the data using pseudo-Voigt and arctangent fit functions. The filled area under the curve is the "fitted area" and corresponds to the white-line intensity. The arctangent fit function is shown as the dashed line and is used to model the continuum step at the current absorption edge. It is notable that the $\mu(E)$ data were normalized so that the continuum step (the height of the high-energy plateau) at the $L_3$-edge is equal to unity for each pressure. Accordingly, the continuum step at the $L_2$-edge has been normalized to half this value (the adopted normalization scheme is the same applied to the iridium-based 5$d$ compounds in [15]). (e) Experimentally observed branching ratios between integrated intensities of the $L_3$ and $L_2$ absorption edges. The solid line at the ratio $I(L_3)/I(L_2) = 2$ corresponds to $\langle \mathbf{L} \cdot \mathbf{S} \rangle = 0$.

Meanwhile, the integrated intensities $I(L_3)$ and $I(L_2)$ of the WL for each pressure were extracted from the raw data, as shown in Fig. 2d. It is known that the ratio $I(L_3)/I(L_2)$ (also called branching ratio (BR)) for 5$d$ transition metal oxides increases with electron occupancy, and for a small electron number the BR is less affected by SOC [18]. However, the BR is related to the ground state expectation value of the angular part of the spin-orbit coupling, $\langle \mathbf{L} \cdot \mathbf{S} \rangle$. Therefore, in the 5$d$ manifold $\langle \mathbf{L} \cdot \mathbf{S} \rangle = n_h(BR - 2)/(BR + 1)$, where $n_h$ is the number of empty holes [19]. As is common, the $\langle \mathbf{L} \cdot \mathbf{S} \rangle$ barely changes within low pressure; therefore, it indicates that the electron-hole density (or electron occupancy) increases (or decreases). These results are consistent with the WL position measurements.

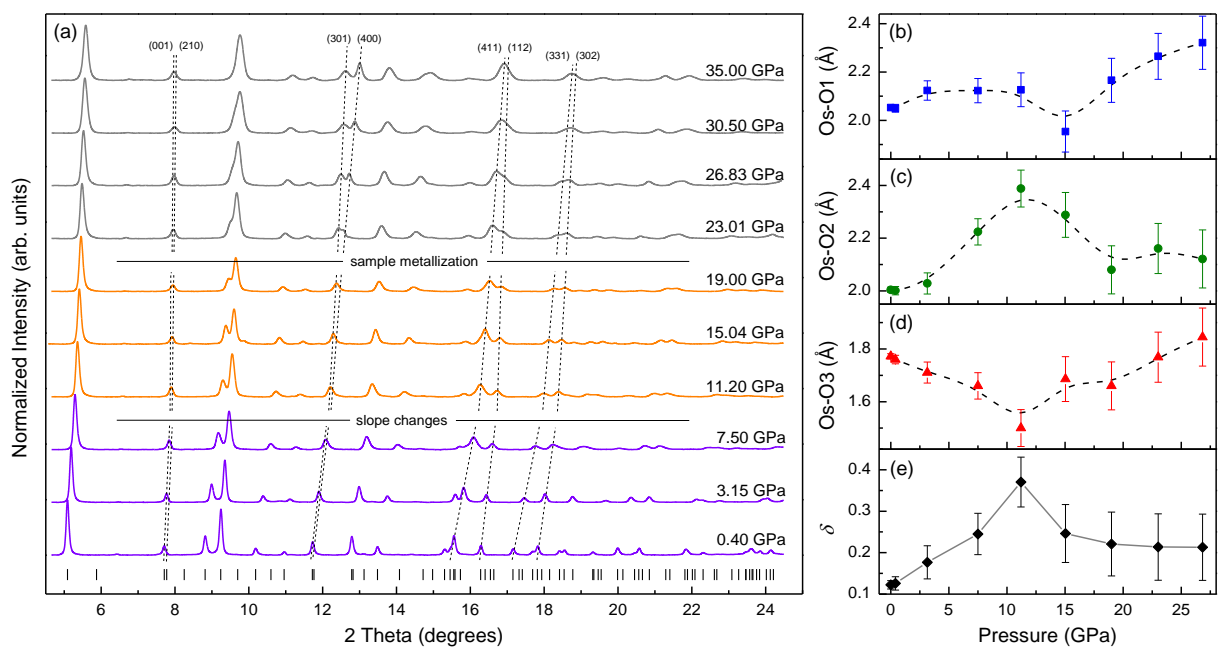

**Figure 3.** Synchrotron X-ray diffraction data collected for $Na_2OsO_4$ powders at room temperature. (a) Evolution of diffraction patterns over sample compression (dashed lines are guides to the eye for randomly selected peaks to highlight structural changes). The XRD peaks move quickly to the higher $2\theta$ angle up to 11 GPa. Further, above 11 GPa, peaks move more slowly, which is different for some peaks, indicating that lattice parameters have different compressibility under pressure. The change in color over pressure denotes the change between different sample states. The positions of the $P\bar{6}2m$ Bragg reflections are marked by vertical bars. (b), (c) and (d) represents variations of the Os-O1, Os-O2, and Os-O3 bond lengths with pressure, respectively. The dashed lines are B-splines used to fit experimental data. (e) The calculated $OsO_6$ octahedral distortion parameter $\delta$ versus pressure.

In addition to the XAS, we also investigated the structural changes at high pressure using diffraction methods; the results are shown in Fig. 3. There are no observable symmetry changes in the diffraction patterns up to 35 GPa in agreement with our optical measurements (for more details see Supporting Information). However, our structural refinement reveals that the $OsO_6$ octahedral distortion remains at high pressure and even increases with pressure up to 11 GPa preserving $S = 0$ state. Although this particular pressure range corresponds with the solidification of neon gas at ambient temperature the transition is consistent with the measurements performed by using other pressure mediums. The evolution of the traced osmium − oxygen bond lengths under high pressure is shown in Figures 3b–d. During compression up to 11 GPa, the atoms O1 and O2 moved away from the osmium giving bond distances of 2.126 Å (Os-O1) and 2.388 Å (Os-O2), respectively. Meanwhile, the O3 atoms were found to move continuously closer to the osmium. The Os-O3 bond length reaches the minimum with a very low value of approximately 1.5 Å at 11 GPa. Further compression above 11 GPa showed opposite behavior of all the atoms, and the difference in the octahedral bonds was always greater than that at the ambient conditions. The distortion in the $OsO_6$ site was characterized at selected pressures by calculating octahedral strain tensor parameter $\delta$ according to the following equation [20]:

$$\delta = \sqrt{\frac{1}{n}\sum_{i=1}^{n}(d_i - d)^2}. \quad (1)$$

Here, $d$ is an average bond length, $d_i$ is an individual bond length, and $n$ is the number of bonds in the octahedra. The parameter $\delta$ indicates the degree of distortion away from the regular octahedron ($\delta = 0$). The larger value of $\delta$, the more distortion of the $OsO_6$ (similarly, the defined equivalent equation – called bond-length distortion [21], can also be used in this case). At ambient pressure, $Na_2OsO_4$ has $\delta = 0.122$ distortion, which is considerably greater than the isostructural $Ca_2IrO_4$, $\delta = 0.038$. However, we found that $\delta$ increases strongly with pressure until 11 GPa. Further increase of pressure forces distortion to drop and proceed with more or less stable value $\delta = 0.2 \pm 0.02$.

Finally, to understand the interplay among the lattice, charge, and orbitals, we performed first-principles calculations based on the density functional theory (DFT). The structural data obtained from the diffraction measurement was used for the calculations. The pressure dependent orbital resolved partial density of states (P-DOS) foremost confirms that $E_g$ is opened at the Fermi level ($E_F$) between $d_{xy}$ and $d_{yz} - d_{zx}$ of the $t_{2g}$ band (see Fig. 4). This indicates that $Na_2OsO_4$ is characteristic of a narrow-band-gap insulator (or semiconductor), rather than a Mott insulator. The low-lying $d_{xy}$ states due to the distorted $OsO_6$ are fully occupied by two electrons, which is also consistent with the non-magnetic configuration. When the pressure increases, the $d_{yz}$, and $d_{zx}$ bands move towards high energy while the $d_{xy}$ band moves towards lower energy; this enlarges the band gap (the extracted activation energy follows this trend similarly), which reaches a maximum at 11 GPa. This result explains why the $Na_2OsO_4$ becomes more insulating at high pressure and the $E_g$ reaches its maximum at 11 GPa. At the same time, the electron occupancy of the $d_{xy}$ band decreases with pressure. Such a reduction of the electron occupancy in the $t_{2g}$ band of Os explains why the WL of $L_3$ edge shifts to higher energy in the XANES spectra and why the BR decreases with pressure reaching a minimum at 11 GPa.

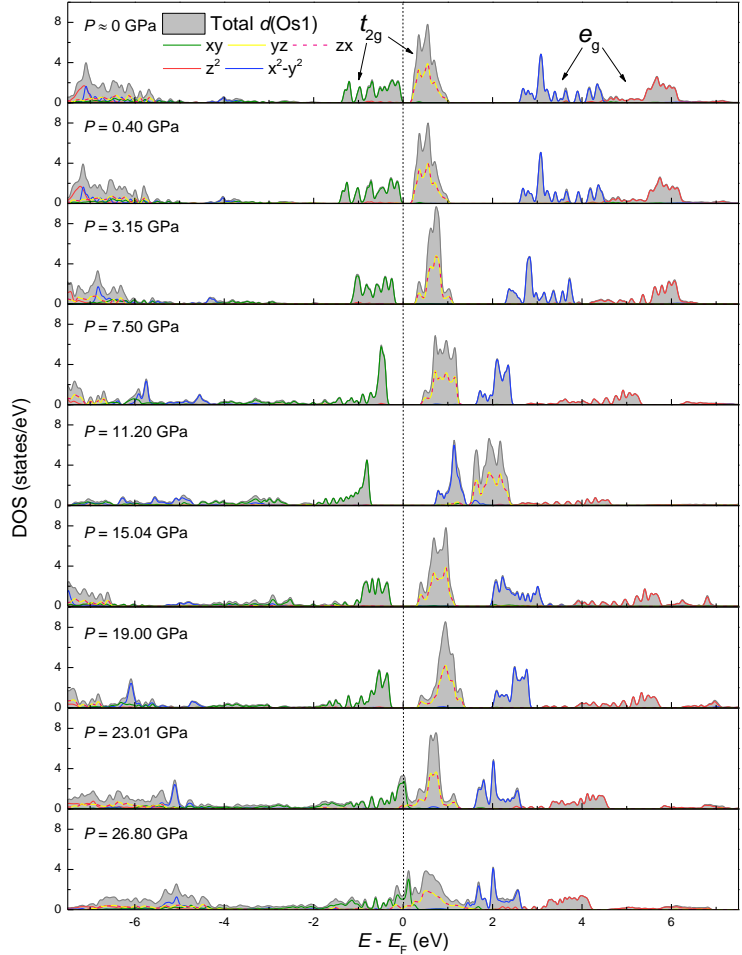

**Figure 4**. The partial density of states for the Os1 atom (with consideration of its local axis) in Na$_2$OsO$_4$. For other Os atoms, the P-DOS are just the same considering their own local axis. The $t_{2g}$ bands at ambient pressure (marked here as $P \approx 0$) originate in the region from -1.4 to 1.05 eV, and $e_g$ bands lay in the conduction band from 2.6 to 7.4 eV. The narrow gap is opened at the top of the valence band in the $t_{2g}$ structure. The $E_F$ is denoted by the dotted line. Note, the P-DOS of the $d_{yz}$ orbital exactly matches the P-DOS of the $d_{zx}$ orbital.

In addition to the change in the $t_{2g}$ band, the $e_g$ band split into $d_{z^2}$ and $d_{x^2-y^2}$. When the pressure increases, the $d_{x^2-y^2}$ band moves towards lower energy while the $d_{z^2}$ moves in the opposite direction. However, when the pressure goes beyond the 11 GPa, all the bands reverse their directions and eventually the $d_{xy}$ and $d_{yz} - d_{zx}$ merge together closing the $E_g$, which results in a metallic state at 23 GPa. This result finally explains why Na$_2$OsO$_4$ becomes more insulating up to 11 GPa, but then gradually turns into a more metallic state and eventually turns into a metal at 23 GPa. The band structure calculations successfully reproduce the observations, basically indicating that the electronic structure of $Na_2OsO_4$ at high pressure is governed by the structure, especially the distortion of $OsO_6$. $Na_2OsO_4$ could have several more features in metallic phase, which are significant for low dimensions, such as the Kondo effect [22], electron fractionalization [23,24], or superconductivity at very low temperature, $T_c < 2$ K [25].

**Experimental and Computational Methods**

The $Na_2OsO_4$ single crystals were obtained using the high-pressure apparatus and route described in [2]. The average crystal size was 0.1–0.3 mm and the shape was needle-like and black in color. Before using the crystals, after washing several times with an ultrasonic bath, XRD test was carried out using Cu $K\alpha$ radiation in SmartLab, RIGAKU to confirm no damage.

A Mao-type symmetric diamond anvil cell (DAC) with 400 μm culet sized anvils was used for the Raman and high-pressure XRD experiments. A stainless steel gasket was precompressed to a 35 μm thickness and a hole of 150 μm was drilled to load the sample, a ruby for pressure determination, and neon gas to serve as a pressure-transmitting medium [26]. The Raman spectra up to ~ 40 GPa were measured on a Renishaw inVia spectrometer with a 488 nm laser wavelength. The data collection time was 90 s and laser power of 15 mW was maintained for each spectrum. The *in situ* high-pressure XRD measurements were carried out in an angle-dispersive mode at beamline 16-BM-D of the Advanced Photon Source (APS), Argonne National Laboratory. The incident monochromatic X-ray beam energy was set to 29.2 keV ($\lambda$ = 0.4246 Å) where the sample-detector distance was 318 mm. The wavelength of the X-ray was periodically calibrated using a $CeO_2$ standard. Diffraction patterns were recorded on a MAR345 image plate and then integrated using DIOPTAS software [27]. Indexing and Rietveld refinements were carried out in EXPO2014 [28] and GSAS-II [29].

High-pressure XAS experiments were performed for osmium by investigating the X-ray absorption near edge structure (XANES) at beamline 20-BM-B of the APS. A panoramic DAC with 400 μm diamonds was used to collect spectra at both the $L_2$ and $L_3$ absorption edges for the Na$_2$OsO$_4$ powders. In order to avoid contamination of the XANES spectra by Bragg peaks from the diamond anvils, XANES measurements were performed in transmission geometry where the X-ray beam goes through a beryllium gasket. The gasket was initially precompressed to ~ 70 μm, and then the whole culet area was drilled and replaced by boron nitride (BN) powder, which was compressed again to make a 60 μm radius hole drilled at the center of the BN insert. The sample, together with a ruby sphere and mineral oil (ACROS Organics™) as a pressure medium, were then all loaded into the prepared hole. The XANES spectra were double-checked by measuring the Na$_2$OsO$_4$ single crystals. In these measurements, we used a symmetric DAC with nanodiamond anvils. A stainless steel gasket was used to make a hole for a sample, ruby, and mineral oil. The collected data was processed and analyzed using programs from the Demeter package [30].

The electronic transport properties under high pressure and low temperature were investigated via the four-probe electrical conductivity method in a DAC made of CuBe alloy. The pressure was generated by a pair of diamonds with a 300 μm diameter culet. A gasket made of stainless steel was pressed from a thickness of 250 μm to 30μm. A hole in the center of the gasket was then drilled with a diameter of 200 μm. Fine cubic BN powder was used to cover the gasket to protect the electrode leads insulated from the metallic gasket. The electrodes were slim Au wires with a diameter of 18μm. A 100μm-diameter center hole in the insulating layer was used as the sample chamber. NaCl powder was used as the pressure-transmitting medium. A thermocouple was mounted near the diamond in the DAC to monitor the exact sample temperature. The measurements were performed using the Mag Lab system, which controls the temperature automatically. The pressure was measured via the ruby fluorescence method at room temperature before and after cooling. It is worth mentioning that in all the experiments pressure value is the average value determined before and after the measurement. In all cases, the error does not exceed ±0.5 GPa.

Electronic structure calculations were performed by the OpenMX software package [31], which is based on the linear combination of pseudo-atomic-orbital basis formalism. The exchange-correlation energy was calculated within the local density approximation (LDA) functional [32]. Spin-orbit coupling was treated in the fully-relativistic j-dependent pseudopotential and non-collinear scheme [33]. The $6 \times 6 \times 12$ Monkhorst-Pack $k$-point grid and the 400 Ry energy cut-off were used for momentum-space and the real-space integration. The theoretical calculation for the band gap verification was also performed for the volume optimized cell following a similar approach applied earlier on the $Na_2OsO_4$ sample [34]. Calculations were double-checked with the full-potential linearized augmented plane wave method as implemented in the Wien2k software along with the Perdew-Burke-Ernzerhof (PBE) parameterized generalized gradient approximation [35]. The $Rk_{max}$ was set to 7 ($R$ is the radius of the smallest muffin-tin sphere and $k_{max}$ is the largest $k$ vector in the plane wave expansion). A mesh of 729 $k$-points in the irreducible part of the Brillouin zone was used. The iteration halted when the criterion for the difference in the eigenvalues was less than 0.0001 between the steps of convergence.

**Acknowledgements**

Portions of this work were performed at HPCAT (Sector 16), XSD (Sector 20), and GSECARS (Sector 13), Advanced Photon Source (APS), Argonne National Laboratory. HPCAT operations are supported by DOE-NNSA's Office of Experimental Sciences. XSD operations are supported by the U.S. Department of Energy (DOE) and the Canadian Light Source (CLS). Use of the COMPRES-GSECARS gas loading system was supported by COMPRES under NSF Cooperative Agreement EAR-1606856 and by GSECARS through NSF grant EAR-1634415 and DOE grant DE-FG02-94ER14466. The APS is a DOE Office of Science User Facility operated for the DOE Office of Science by Argonne National Laboratory under Contract No. DE-AC02-06CH11357. Y.D and H.-k. M. acknowledges the support from DOE-BES under Award No. DE-FG02-99ER45775 and NSFC Grant No. U1530402. H.Y., M.Y.J., and M.J.H. acknowledges the Basic Science Research Program through the National Research Foundation of Korea (NRF) funded by the Ministry of Education (2018R1A2B2005204). This work was supported by National Key R&D Program of China 2018YFA0305703 and Science Challenge Project, No TZ2016001. The work in Japan was supported in part by JSPS KAKENHI Grant Numbers JP15K14133 and JP16H04501.

# Supplementary data

In the current crystal setting, $OsO_6$ is not aligned with any axis; therefore, we recalculated the partial density of states (P-DOS) of the Os atoms with consideration of their local axes for pressures up to 26.8 GPa (Fig. 4 in the main text). The $Na_2OsO_4$ has a large DOS at the Fermi level ($E_F$) due to Os $d$ states. In the oxide compounds the electron correlations are larger, and the $d$ states are more localized, i.e., they develop larger density of states at $E_F$. The total DOS of $Na_2OsO_4$ calculated by using the LDA potential is shown here in Fig. S1. As discussed in the main text, the compression expands the energy gap ($E_g$) greatly up to 11 GPa and then shrinks it with more compression leading to a closure at 23 GPa.

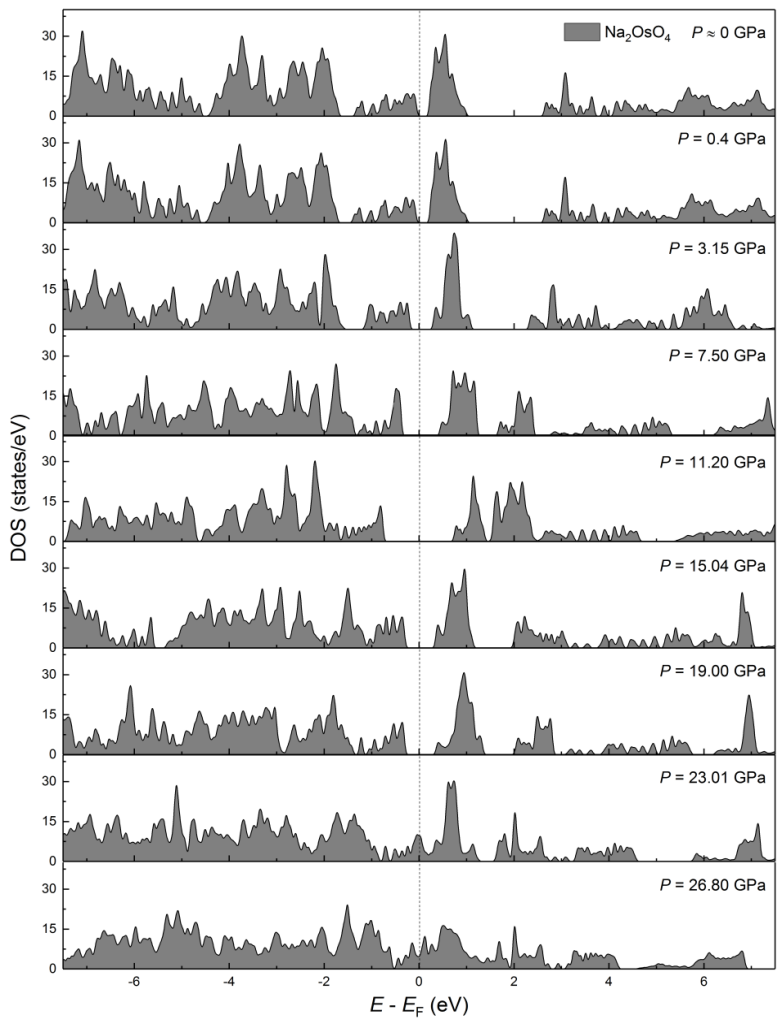

**Supplementary Figure 1.** The total density of states for $Na_2OsO_4$ in the energy range from -7.5 to 7.5 eV. The $E_F$ is denoted by the dotted line.

Within all the pressures up to the 23 GPa, $Na_2OsO_4$ is a semiconductor with an indirect-gap, because the closest states above and beneath the band gap do not have the same wave vector *k* value. The electronic band structures indicating the band gap type at ambient pressure and 11.2 GPa are given in Figs. S2b and S2c, respectively. The calculation was carried out following the general path ($\Gamma$–M–K–$\Gamma$–A–L–H–A|L–M|K–H) through the Brillouin zone of the hexagonal lattice (Fig. S2a) [1].

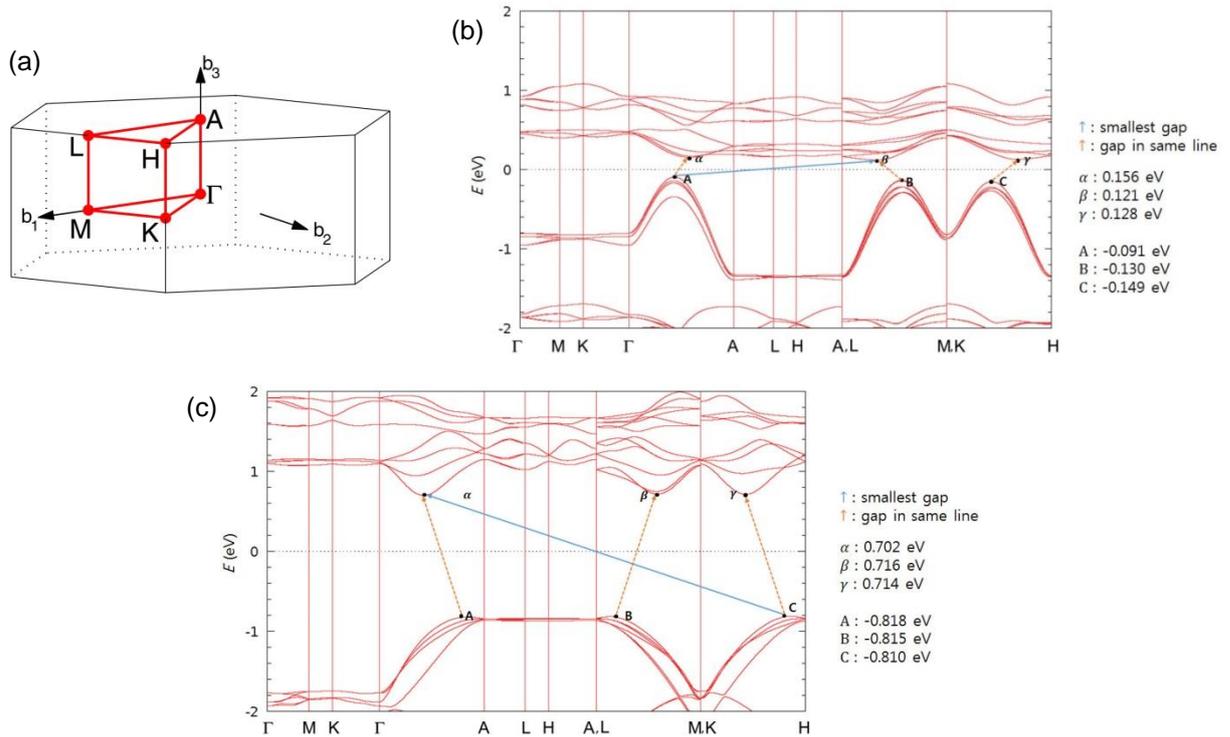

**Supplementary Figure 2.** Calculation of the $Na_2OsO_4$ band structure. (a) The first Brillouin zone of the hexagonal lattice. (b) The band structures of $Na_2OsO_4$ at ambient pressure and (c) at 11.2 GPa. The $E_F$ is denoted by the dotted line. The blue lines indicate the smallest gap, which is obviously an indirect type for both pressures.

The orbital resolved P-DOS confirms earlier predictions that the Os $t_{2g}$ bands are split at $E_F$ by the $E_g$ and the $e_g$ bands are well separated in the energy by the crystal electric field (CEF). The competition between the axial compression and CEF under high pressure is evident. The axial compression weakens CEF from a maximum of 5.3 eV at ambient pressure to about 3.4 eV at 11 GPa (see the imitated orbital energy diagrams presented in Fig. S3). This is unusual because broadened bands typically increase CEF under high pressure. Our theoretical results also show that at 11 GPa, the $d_{x^2-y^2}$ orbital lowers in energy and interrupts the $t_{2g}$ orbitals. From the energy diagrams, one can see that the number of holes in the $t_{2g}$ bands is definitely increased in agreement with our XANES results.

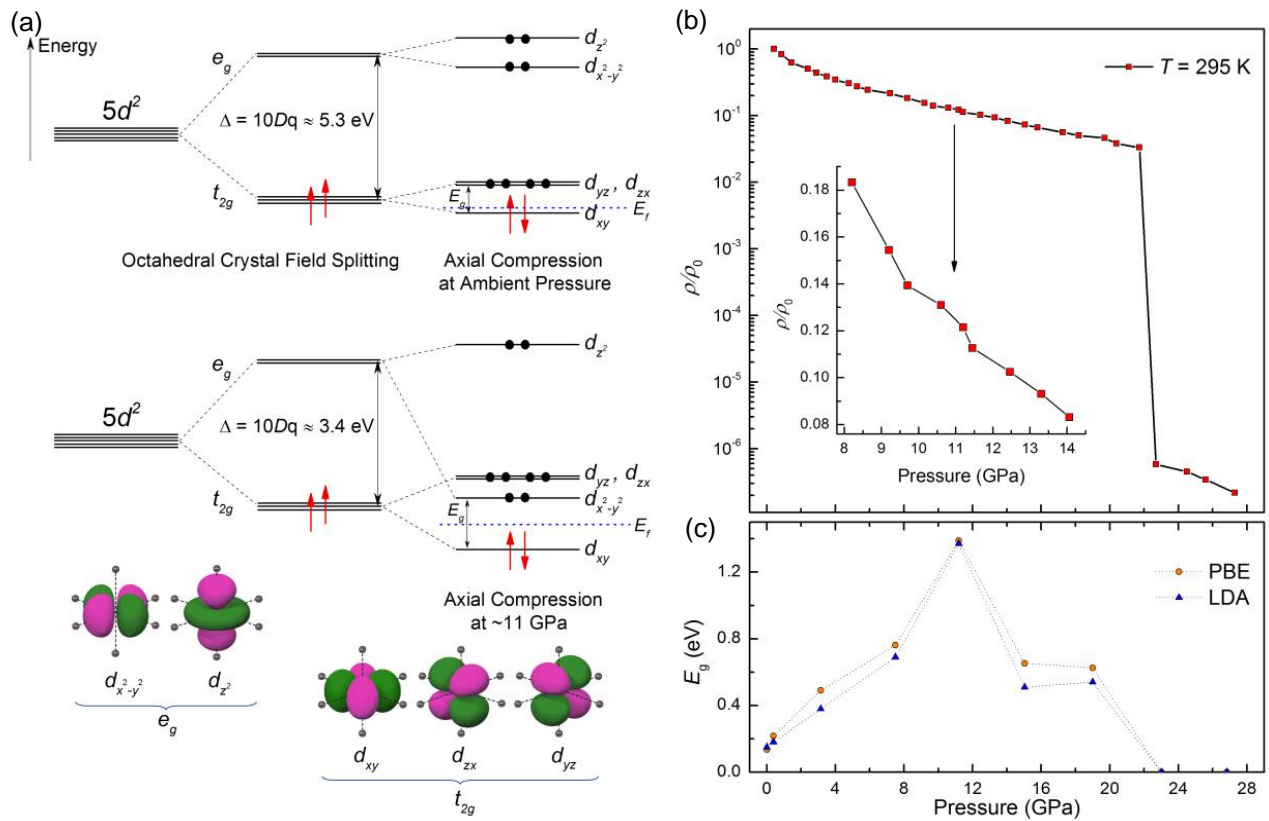

**Supplementary Figure 3.** A comparison between the theoretical calculation results and electrical resistivity measurements. (a) Schematic view of the $5d$ electronic states of $Na_2OsO_4$ $d$-orbital splitting in the octahedral field at different pressures. (b) Normalized electrical resistivity $\rho$ during sample compression. Here, $\rho_0$ is the resistivity at ambient pressure. The sharp drop in the resistivity curve at 22.7 GPa is highlighted using a logarithmic scale and indicates the semiconductor-to-metal transition. The inset shows changes near the highest $OsO_6$ distortion point that was observed using XRD. (c) The theoretical DFT calculations with different potentials show the variation of the $E_g$ during sample compression. The $E_g$ is closed at 23 GPa, in agreement with electrical transport measurements.

In $Na_2OsO_4$, we followed the $E_g$ of the indirect type, which surprisingly increases up to ~ 11 GPa despite the fact that the resistivity decreases continuously (Figs. S3b, c). The pressure dependent electrical resistivity ($\rho$) measured at room temperature up to ~ 23 GPa demonstrated an exponential decrease with increasing pressure; in the logarithmic scale, it shows almost linear behavior (see Fig. S3b). The inset of Fig. S3b highlights a small hump and slope change of the resistivity data at around 11 GPa, where the BR and Os valence have extremum. As our DFT calculation shows, the gap starts to decrease only above 11.5 GPa and closes at 23 GPa confirming the semiconductor-to-metal transition seen as a sharp drop in the $\rho(P)$ curve. In the semi-log plot (Fig. S4) the difference between semiconducting and metallic states is even more vivid. Since these curves have strong curvature the activation energy, which is presented in the main text (Fig. 1b), was extracted from the higher temperature region marked by the dashed line (see Fig. S4).

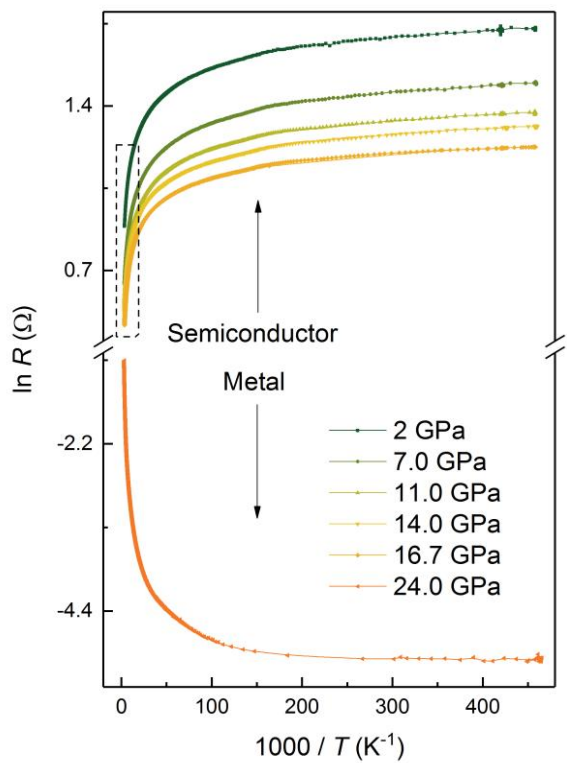

**Supplementary Figure 4.** Arrhenius plot of the electrical resistance $R$ at fixed pressures. The dashed line indicates the area for which the activation energy was extracted.

The X-ray absorption near edge spectroscopy (XANES) measurement at the Os $L_3$ and $L_2$ edges were performed for $Na_2OsO_4$ single crystals. The spectrum at the ambient conditions for the selected sample is given in Fig. S5. The ratio of the integrated intensities for the $L_3$:$L_2$ was found to be 2.36:1. The deviation of the white-line position associated with the Os charge is presented in Fig. S6.

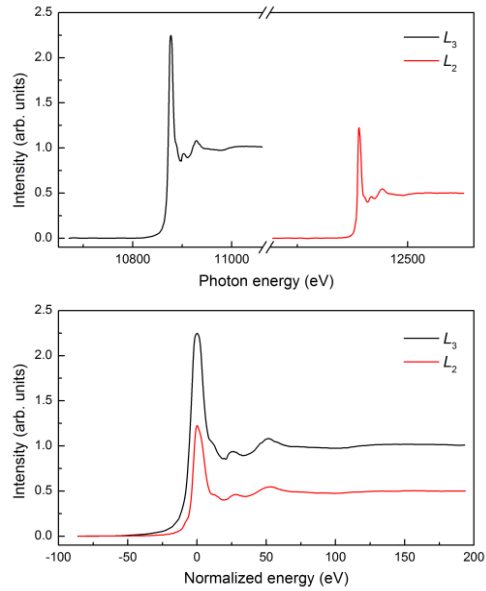

**Supplementary Figure 5.** The XANES spectra of the Os $L_3$ and $L_2$ edges measured in $Na_2OsO_4$ at ambient pressure and room temperature. Note that the intensity was normalized so that the continuum step at the $L_3$-edge is equal to unity for each pressure. Accordingly, the continuum step at the $L_2$-edge was normalized to half this value.

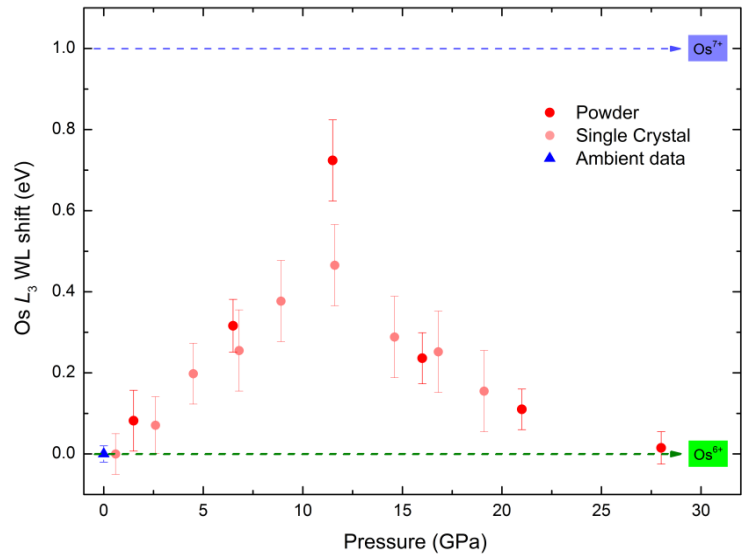

**Supplementary Figure 6.** The Os $L_3$-edge shift indicating Os charge changes in $Na_2OsO_4$ during compression. The maximum deviation of $\Delta = 0.724$ eV is achieved at 11.2 GPa. Here, the dashed line at 0 eV represents the $Os^{6+}$ state, while the dashed line at 1 eV is the hypothetical $Os^{7+}$ valence state for $Na_2OsO_4$.

The $Na_2OsO_4$ under compression manifested as a soft material where the lattice parameters have different compressibility (see Fig. S7). In this system, the $OsO_6$ octahedron is situated so that the *c*-axis is sustained by two pairs of similar lengths of Os-O1 and Os-O2 bonds, while 2x Os-O3 (which are much shorter) rest in the *a-b* plane. The strong $OsO_6$ distortion under high-pressure suggests that $Na_2OsO_4$ remains in a nonmagnetic $S = 0$ state for all the pressures applied, except the metallic state above 23 GPa, where the $d$ orbitals of Os start to mix with each other. More details on $OsO_6$ octahedral are given in Table S1. The refined lattice parameters and atomic positions for $Na_2OsO_4$ at different pressures are given in Table S2.

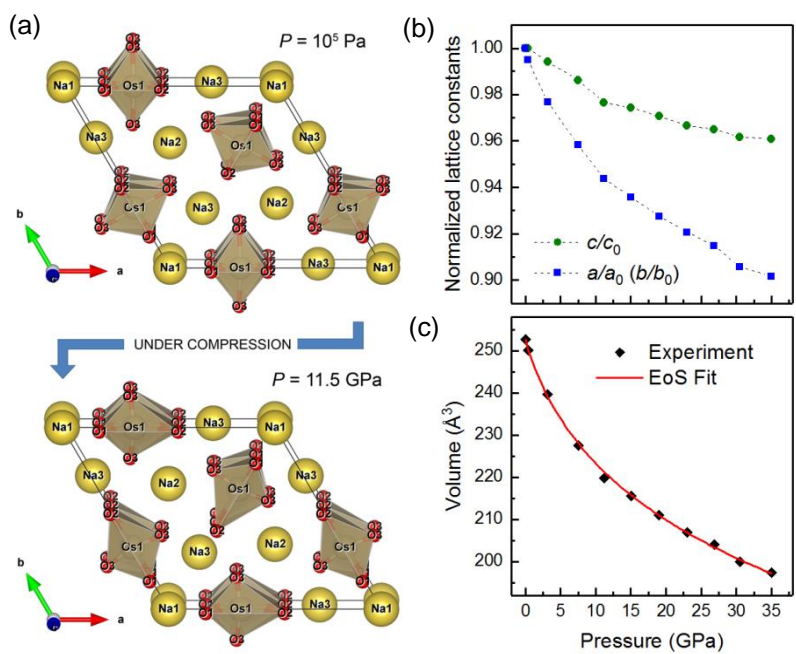

**Supplementary Figure 6.** The structural view of $Na_2OsO_4$ and its lattice change during compression. (a) Crystal structural view at ambient and at 11.5 GPa pressure. The oxygen-coordinated Os atoms are indicated as the $OsO_6$ octahedral. Thick solid lines indicate the hexagonal unit cell. (b) Pressure dependence of the $a/a_0$ ($b/b_0$) and $c/c_0$ ratios. Here, $a_0$ and $c_0$ are the lattice constants at ambient pressure. (c) Pressure dependence of the unit-cell volume. A solid red curve is the calculated third-order Birch−Murnaghan equation of state (EoS) fitted to the experimental data using the program EoSFit [2]. The best-fit yielded the bulk modulus, $K_0 = 42.8$ GPa, and its derivative $K'_0 = 11.5$. The bulk modulus confirms the fact that $Na_2OsO_4$ is soft under high-pressure; its value is two or more times smaller than those observed in different oxides with an octahedrally coordinated metal center [3].

The Raman spectra were tested in the frequency range from 100 to 3200 cm$^{-1}$. The strongest mode, possibly the stretching mode of $OsO_6$ octahedral (further studies are needed), was selected for the investigation under high pressure (Fig. S8). The others were too weak to be qualitatively investigated under high pressure. Nevertheless, during compression, we did not observe any sudden peak appearance or disappearance, which could be a signal of structural change. For all three samples that we took, the main mode showed a consistent frequency shift with increasing pressure. The width and height of this mode indicated changes attributed to the phenomena described in the article. The evidence for the first critical point at 11 GPa is a clear slope change in the course of the peak width. The second critical point at 23 GPa manifested in gradual peak broadening-disappearance together with a decrease in its intensity; typical of a sample metallization process at high pressures.

**Supplementary Table 1.** The angles in the OsO$_6$ at different pressures.

| Pressure, GPa | Angles, deg. | | | | | |
|---|---|---|---|---|---|---|
| | O1-Os-O2 | O1-Os-O3 | O2-Os-O3 | O1-Os-O1 | O2-Os-O2 | O3-Os-O3 |
| **0.0001** | 77.68 | 88.98 | 90.99 | 100.69 | 103.94 | 176.80 |
| **0.40** | 77.42 | 89.00 | 90.96 | 100.98 | 104.19 | 176.86 |
| **3.15** | 81.00 | 93.00 | 87.20 | 95.10 | 102.90 | 171.10 |
| **7.50** | 88.60 | 88.66 | 91.50 | 94.23 | 88.70 | 175.86 |
| **11.20** | 93.43 | 86.05 | 94.45 | 92.83 | 80.34 | 168.46 |
| **15.04** | 85.83 | 88.05 | 92.44 | 103.93 | 84.54 | 173.46 |
| **19.00** | 87.50 | 93.60 | 86.54 | 90.13 | 94.95 | 169.85 |
| **23.01** | 92.74 | 96.66 | 83.73 | 84.73 | 89.85 | 162.25 |
| **26.83** | 93.74 | 100.86 | 80.03 | 81.43 | 91.26 | 151.46 |

**Supplementary Table 2.** The refinement results of the lattice parameters and atomic positions.

| Site | Wyckoff position | 0.0001 GPa | | | 0.40 GPa | | | 3.15 GPa | | |
|---|---|---|---|---|---|---|---|---|---|---|
| | | $a$ = 9.61333 Å, $c$ = 3.15673 Å, $V$ = 252.640 Å$^3$ | | | $a$ = 9.56820 Å, $c$ = 3.15820 Å, $V$ = 250.397 Å$^3$ | | | $a$ = 9.37778 Å, $c$ = 3.13270 Å, $V$ = 238.589 Å$^3$ | | |
| | | $x$ | $y$ | $z$ | $x$ | $y$ | $z$ | $x$ | $y$ | $z$ |
| **Os** | 3f | 0.32963 | 0.00000 | 0.00000 | 0.32960 | 0.00000 | 0.00000 | 0.33290 | 0.00000 | 0.00000 |
| **Na1** | 1a | 0.00000 | 0.00000 | 0.00000 | 0.00000 | 0.00000 | 0.00000 | 0.00000 | 0.00000 | 0.00000 |
| **Na2** | 2d | 0.33333 | 0.66667 | 0.50000 | 0.33333 | 0.66667 | 0.50000 | 0.33333 | 0.66667 | 0.50000 |
| **Na3** | 3g | 0.69434 | 0.00000 | 0.50000 | 0.69430 | 0.00000 | 0.50000 | 0.69800 | 0.00000 | 0.50000 |
| **O1** | 3g | 0.19355 | 0.00000 | 0.50000 | 0.19350 | 0.00000 | 0.50000 | 0.18000 | 0.00000 | 0.50000 |
| **O2** | 3g | 0.45805 | 0.00000 | 0.50000 | 0.45810 | 0.00000 | 0.50000 | 0.46600 | 0.00000 | 0.50000 |
| **O3** | 6j | 0.43085 | 0.21274 | 0.00000 | 0.43090 | 0.21270 | 0.00000 | 0.45200 | 0.21000 | 0.00000 |
| | | 7.50 GPa | | | 11.20 GPa | | | 15.04 GPa | | |
| | | $a$ = 9.20128 Å, $c$ = 3.10910 Å, $V$ = 227.96 Å$^3$ | | | $a$ = 9.06642 Å, $c$ = 3.08118 Å, $V$ = 219.341 Å$^3$ | | | $a$ = 8.99983 Å, $c$ = 3.07713 Å, $V$ = 215.846 Å$^3$ | | |
| | | $x$ | $y$ | $z$ | $x$ | $y$ | $z$ | $x$ | $y$ | $z$ |
| **Os** | 3f | 0.33510 | 0.00000 | 0.00000 | 0.33970 | 0.00000 | 0.00000 | 0.33680 | 0.00000 | 0.00000 |
| **Na1** | 1a | 0.00000 | 0.00000 | 0.00000 | 0.00000 | 0.00000 | 0.00000 | 0.00000 | 0.00000 | 0.00000 |
| **Na2** | 2d | 0.33333 | 0.66667 | 0.50000 | 0.33333 | 0.66667 | 0.50000 | 0.33333 | 0.66667 | 0.50000 |
| **Na3** | 3g | 0.70900 | 0.00000 | 0.50000 | 0.28900 | 0.28900 | 0.50000 | 0.28100 | 0.28100 | 0.50000 |
| **O1** | 3g | 0.17800 | 0.00000 | 0.50000 | 0.17800 | 0.00000 | 0.50000 | 0.20300 | 0.00000 | 0.50000 |
| **O2** | 3g | 0.50800 | 0.00000 | 0.50000 | 0.45900 | 0.45900 | 0.50000 | 0.47500 | 0.47500 | 0.50000 |
| **O3** | 6j | 0.43300 | 0.20900 | 0.00000 | 0.41800 | 0.19000 | 0.00000 | 0.43400 | 0.21600 | 0.00000 |
| | | 19.00 GPa | | | 23.01 GPa | | | 26.83 GPa | | |
| | | $a$ = 8.92474 Å, $c$ = 3.06655 Å, $V$ = 211.53 Å$^3$ | | | $a$ = 8.8504 Å, $c$ = 3.05114 Å, $V$ = 206.975 Å$^3$ | | | $a$ = 8.79327 Å, $c$ = 3.04534 Å, $V$ = 203.923 Å$^3$ | | |
| | | $x$ | $y$ | $z$ | $x$ | $y$ | $z$ | $x$ | $y$ | $z$ |
| **Os** | 3f | 0.32840 | 0.00000 | 0.00000 | 0.32610 | 0.00000 | 0.00000 | 0.32530 | 0.00000 | 0.00000 |
| **Na1** | 1a | 0.00000 | 0.00000 | 0.00000 | 0.00000 | 0.00000 | 0.00000 | 0.00000 | 0.00000 | 0.00000 |
| **Na2** | 2d | 0.33333 | 0.66667 | 0.50000 | 0.33333 | 0.66667 | 0.50000 | 0.33333 | 0.66667 | 0.50000 |
| **Na3** | 3g | 0.28100 | 0.28100 | 0.50000 | 0.28600 | 0.28600 | 0.50000 | 0.27100 | 0.27100 | 0.50000 |
| **O1** | 3g | 0.15700 | 0.00000 | 0.50000 | 0.13700 | 0.00000 | 0.50000 | 0.12400 | 0.00000 | 0.50000 |
| **O2** | 3g | 1.00000 | 0.48600 | 0.50000 | 1.00000 | 0.49900 | 0.50000 | 1.00000 | 0.49500 | 0.50000 |
| **O3** | 6j | 0.45200 | 0.21400 | 0.00000 | 0.47100 | 0.22800 | 0.00000 | 0.49400 | 0.23400 | 0.00000 |

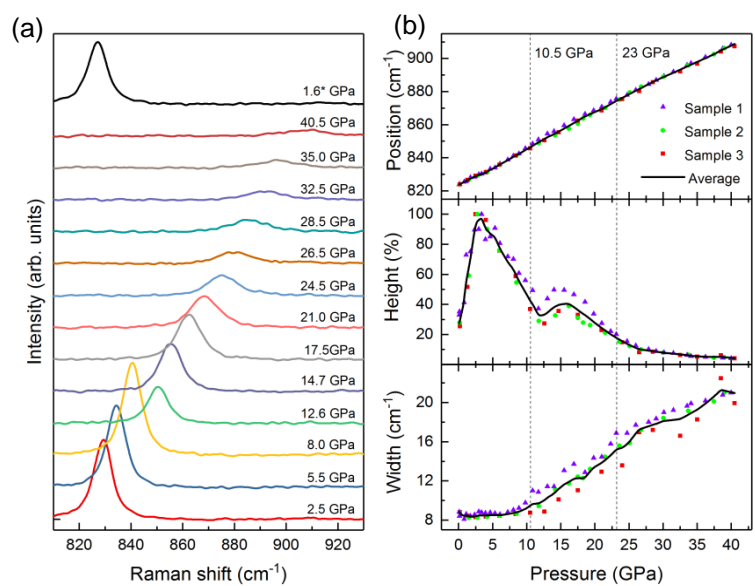

**Supplementary Figure 8.** Pressure-dependent evolution of the strongest mode of $Na_2OsO_4$ samples at room temperature. (a) Selected Raman spectra for the "Sample 3" at various pressures during compression. The release spectrum obtained at 1.6 GPa is marked with an asterisk. (b) The data points indicating the peak position, height and width repeated for three of the $Na_2OsO_4$ samples. The black solid curves represent the averaged data. Note that the laser power and the measuring time were kept constant for all spectra.

## Supplementary References